\documentstyle[a4,12pt]{article}
\begin{document}
\begin{titlepage}
\renewcommand{\thefootnote}{\fnsymbol{footnote}}
\rightline{DAMTP-R93/9}
\vspace{0.6in}
\LARGE
\center{On the Energy Levels of the Hydrogen Atom}
\Large
\vspace{1.2in}
\center{C.J. Fewster\footnote{E-mail address:
C.J.Fewster@amtp.cam.ac.uk}} \vspace{0.2in}
\large
\center{\em Department of Applied Mathematics and Theoretical Physics,
\\ University of Cambridge,
\\  Silver Street, Cambridge CB3 9EW, U.K.}
\vspace{0.3in}
\center{May 11, 1993}
\vspace{1in}
\begin{abstract}
We re-examine the  justification for the imposition of regular
boundary conditions on the wavefunction at the Coulomb singularity in
the treatment of the hydrogen atom in non-relativistic quantum
mechanics. We show that the issue of the correct boundary conditions
is not independent of the physical structure of the proton. Under the
physically reasonable assumption that the finite size and structure
of the proton can be represented as a positive correction to the
Coulomb potential, we give a justification for the regular boundary
condition, which, in contrast to the usual treatments, is physically
motivated and mathematically rigorous. We also describe how irregular
boundary conditions can be used to model non-positive corrections to
the Coulomb potential.
\end{abstract}
\setcounter{footnote}{0}
\renewcommand{\thefootnote}{\arabic{footnote}}
\end{titlepage}

\input mssymb
\newcommand{\Avec}{{\bf A}}
\newcommand{\Bvec}{{\bf B}}
\newcommand{\Lvec}{{\bf L}}
\newcommand{\Rvec}{{\bf R}}
\newcommand{\zvec}{{\bf \hat{z}}}
\newcommand{\nvec}{{\bf n}}
\newcommand{\xvec}{{\bf x}}
\newcommand{\pvec}{{\bf p}}
\newcommand{\rvec}{{\bf r}}
\newcommand{\thvec}{\bf \hat{\hbox{\boldmath $\theta$}}}
\newcommand{\HH}{{\cal H}}
\newcommand{\CC}{{\Bbb C}}
\newcommand{\DD}{{\cal D}}
\newcommand{\NN}{{\Bbb N}}
\newcommand{\RR}{{\Bbb R}}
\newcommand{\Coinf}{C_0^{\infty}({\Bbb R}^3\backslash\lbrace 0\rbrace)}
\newcommand{\Seq}{\Sigma}
\newcommand{\Cl}{{\cal V}}
\newcommand{\Hi}{H_{\rm ideal}}
\newcommand{\Hia}{H^*_{\rm ideal}}
\newcommand{\Emb}{{\cal J}}
\newcommand{\inner}[2]{\langle #1 \mid #2 \rangle}
\newtheorem{Thm}{Theorem}[section]
\newtheorem{Lem}[Thm]{Lemma}
\newtheorem{Def}[Thm]{Definition}
\newtheorem{Prop}[Thm]{Proposition}
\newtheorem{Cor}[Thm]{Corollary}
\newtheorem{Rcp}{Theorem}
\newenvironment{Recap}[1]{
\renewcommand{\theRcp}{{#1}}
\begin{Rcp}}{\end{Rcp}}
\renewcommand{\theequation}{\thesection.\arabic{equation}}

\newcommand{\sect}[1]{\section{#1}\setcounter{equation}{0}}

\sect{Introduction}

One of the unsatisfactory features of Old Quantum Theory was the
primacy of the concept of `quantum number', which does not arise from
more basic physical principles thereby leaving a certain mystery as
to the true origin of discreteness in quantum theory and the
distinguished nature of the positive integers. It was to remove this
limitation that Schr\"{o}dinger embarked upon a programme of
re-casting quantisation as an eigenvalue problem, studying the
hydrogen atom as his first example \cite{Schr}.

Of course, eigenvalue problems such as those encountered in
quantum theory require the specification of boundary conditions in
order to be well-posed. In the hydrogen atom case, there is a
particular issue about what boundary conditions should be imposed on
the wavefunction at the Coulomb singularity. This was understood by
Schr\"{o}dinger, who imposed that the wavefunction be finite. With
this condition, the negative energy eigenvalues of the idealised
Coulomb Hamiltonian precisely replicate the Bohr levels, a success
which confirmed the validity of the new approach. As Schr\"{o}dinger
wrote \cite{Schr}
\begin{quote}
The essential thing seems to me to be, that the postulation of
``whole numbers'' no longer enters into the quantum rules
mysteriously, but that we have traced the matter a step further back,
and found the ``integralness'' to have its origin in the finiteness
and single-valuedness of [the wavefunction]
\end{quote}

But is finiteness of the wavefunction an axiom of quantum mechanics?
Although this issue is often considered in quantum mechanics
textbooks, there does not appear to be a satisfactory treatment in
the literature -- see \cite{FK} for a discussion of the usual
treatments. In the present paper, we examine this problem in detail
for the particular case of the hydrogen atom. One of our main aims
will be to provide a physically motivated and mathematically rigorous
justification for the condition that the wavefunction should be
finite at the Coulomb singularity. We shall also consider the
interpretation of other possible boundary conditions as models for
nuclear structures differing greatly from that of the physical proton.

Our general methodology, which is discussed at length in \cite{FK},
starts from the distinction introduced in \cite{KF} between {\em true
problems} and {\em idealised problems}.  In a true problem,
interactions are modelled by a smooth potential in the
Schr\"{o}dinger equation, reflecting our expectation that only smooth
potentials occur in nature. In the case of the hydrogen atom, one can
identify a class of true problems consisting of models in which the
finite size and structure of the proton is modelled by a smooth
potential which deviates from the Coulomb form within the nuclear
radius, `rounding off' the Coulomb singularity. For simplicity, we
shall mainly consider spherically symmetric potentials of this form,
but we make no other restriction on our class of true problems. For
true problems, we will take finiteness and smoothness of the
wavefunction to be axiomatic for quantum mechanics.

Idealised problems represent limiting cases of true problems.
As such, they can possess distinguished points at which some of the
structure of a more fundamental true problem has been simplified. In
the case of the hydrogen atom, the Coulomb singularity is a
distinguished point, for it represents the physical proton of the
true problem. Note that the idealised problem need not be singular
at such points. For example, a true problem in which the potential
changes steeply but smoothly from one value to another could be
idealised by a potential step. One might idealise a true problem
with a potential compactly supported within a small neighbourhood of
the origin, by an idealised problem with no potential at all, but
with the origin as a distinguished point (see \cite{KF}, where this
case is treated in depth). At such distinguished points in idealised
problems, boundary conditions can only be physically justified by
reference to the true problem one intends the idealisation to
represent.

Accordingly, in Section 2, we determine the class of allowed
idealised problems for the case in hand by considering limits (in
the strong resolvent sense \cite{RSi}) of sequences of Hamiltonians
for true hydrogen atom problems whose `nuclear radii' tend to zero.
We thereby determine a 1-parameter family of idealised problems with
different boundary conditions at the Coulomb singularity, all of
which correspond to the limiting behaviour of true problems from our
class. The spectra of these idealised problems can differ markedly
in the $S$-wave, although for higher angular momenta, all
idealisations yield the Bohr levels.

Clearly, in order to justify the choice of regular boundary condition
at the origin (the unique choice leading to the Bohr levels in the
$S$-wave) we must make further restrictions on our class of true
problems. In Section 3, we make the physically reasonable restriction
to true problems whose potentials represent {\em positive}
corrections to the Coulomb potential localised within a small nuclear
radius. Under these assumptions, we prove that the true problem has
energy levels which differ only slightly from those of the idealised
system with the regular boundary condition (i.e. the Bohr levels).
There are, of course, well known estimates from first order
perturbation theory of the finite size corrections to the Bohr levels
under the further assumption that the nucleus is a smooth
positive charge distribution. However, our treatment will cover a
wide class of nuclear models in a non-perturbative fashion. We also
note that this argument is not usually given as the `correct'
understanding of the point in question. We will also show that there
exist true problems with arbitrarily small nuclear radius (with
non-positive corrections to the Coulomb potential) whose $S$-wave
energy levels differ greatly from the Bohr levels.

We also provide a second justification by considering limits in the
sense of strong resolvent convergence of sequences of true problem
Hamiltonians whose nuclear radii tend to zero, intepreting this as a
means of `regularising' the point charge. By definition, in the
general case, the class of possible self-adjoint limits is precisely
the class of idealised problems, which (as we mentioned above) is a
1-parameter family of operators with different boundary conditions at
the origin. However, restricting to regularisations which can be
modelled by  positive, compactly supported corrections to the Coulomb
potential, we will prove that all such regularisation schemes
converge rigorously to the idealised Hamiltonian with regular
boundary conditions.

In Section 4, we turn to the intepretation of irregular
boundary conditions at the Coulomb singularity, by developing
a scattering  length formalism for systematically
matching any given true problem to its `best fit' idealisation. This
formalism also plays a r\^ole in the proof of our convergence
theorems mentioned above. In Section 5, we examine the situation for
angular momenta $\ell\ge 1$. We find that the Bohr spectrum in these
sectors is more stable against finite size effects than in the
$S$-wave. Section 6 contains the rigorous proofs of some of the
results stated in earlier sections.

One of the motivations for this work was the recent study undertaken
by Kay and the author of the r\^{o}le of model dependence in systems
which interact with small objects \cite{KF}, and in particular the
`principle of sensitivity' introduced in \cite{KF} (see also
\cite{KS}) which relates the range of possible large scale (low
energy) behaviour of a class of true problems  to the range of
possible behaviour exhibited by the corresponding class of  idealised
problems. We conclude in Section 7, by discussing the relation of our
present results to the principle of sensitivity, with which we find
good agreement.

\sect{True and Idealised Problems}

The Hamiltonian corresponding to the idealised hydrogen problem is
given as a differential operator by
\begin{equation}
\Hi = -\triangle +\frac{\gamma}{r} ,\label{eq:Hi}
\end{equation}
where we employ units in with $\hbar=4\pi\epsilon_o=1$ and in which
the reduced mass of the electron is set to $\frac{1}{2}$ ($\gamma$ is
negative). The Bohr radius is thus given by $a_0=2/|\gamma|$. Our
Hilbert space of wavefunctions is $\HH=L^2(\RR^3,d^3\rvec)$ with inner
product denoted $\inner{\cdot}{\cdot}$. To define $\Hi$ rigorously
on $\HH$, we must specify its domain, which inevitably begs the
question of the appropriate boundary conditions at the Coulomb
singularity. In order to circumvent this, we consider the class of
true problems, in which the singularity is rounded off within
some nuclear radius, and for which we assume finiteness of the
wavefunction as an axiom. We then {\em define} the class of idealised
Hamiltonians to consist of the self-adjoint limits in the strong
resolvent sense of sequences of true problem Hamiltonians
whose nuclear radii tend to zero.

Let $\Cl$ be the class of all measurable real-valued functions $V(r)$
on $\RR^+$ compactly supported within some radius of the origin such
that $\rvec\mapsto \gamma |\rvec|^{-1}+V(|\rvec|)$ is smooth  on
$\RR^3$. Then the class of true problems consists of all Hamiltonians
defined as the closure of an operator of form
\begin{equation}
-\triangle +\frac{\gamma}{r}+V(r) \quad {\rm on}~ C_0^\infty(\RR^3)
\subset \HH \label{eq:Htrue}
\end{equation}
for $V\in\Cl$. We refer to the radius of support of $V$ as the {\em
nuclear radius}. Operators of form~(\ref{eq:Htrue})
are automatically essentially self-adjoint as a consequence of the
Kato-Rellich theorem by hypothesis on $\Cl$ (see Theorem X.15 in
\cite{RSii}). This definition encodes two desirable features of true
problems: firstly that only smooth potentials occur in nature, and
secondly that the wavefunction is everywhere finite. The latter
follows from our choice of domain, which has already injected regular
boundary conditions at the origin.

We now define the class $\Seq$ of all sequences of true problem
Hamiltonians $H_n=-\triangle+\gamma/r+V_n(r)$ whose nuclear radii
(i.e. the radii of the supports of the $V_n$ about the origin) tend
to zero. Our idealisations are the self-adjoint limits of such
sequences in the {\em strong resolvent sense}.

The following result, which we prove in Section 6, classifies
the idealised Hamiltonians as the self-adjoint extensions of the
symmetric operator $\Hi$ on the domain $\Coinf$.
\begin{Thm}
\label{Th:seq}
 Let $\Hi=-\triangle+\gamma/r$ on $\Coinf$. (a) Let $\{H_n\}\in\Seq$.
If $\{H_n\}$ has a self-adjoint limit $H$ in the strong resolvent
sense, then $H$ is a self-adjoint extension of $\Hi$. (b)
Furthermore, all self-adjoint extensions of $\Hi$ arise as the strong
resolvent limits of sequences in $\Seq$.
\end{Thm}

We note that these extensions have also been exhibited as limits of
sequences of self-adjoint operators by Albeverio and co-workers
\cite{Alb}, where however, the convergence is in the norm resolvent
sense. We motivate our choice of convergence by two observations.
Firstly, the strong resolvent convergence of a sequence of
self-adjoint Hamiltonians to a self-adjoint limit is equivalent
(Trotter's Theorem \cite{RSi}) to the statement that
\begin{equation}
 e^{iH_nt}\psi\rightarrow
e^{iHt}\psi \quad {\rm for~all}~t,~{\rm and~all}~\psi\in\HH.
\end{equation}
Thus strong resolvent convergence is a natural candidate for the
notion of dynamical convergence. Under certain circumstances, it can
be shown that if $H_n\rightarrow H$ in the strong resolvent sense,
then the corresponding M{\o}ller wave operators $\Omega^\pm(H_n,H_o)$
converge strongly to $\Omega^\pm(H,H_o)$, where $H_o$ is a suitable
comparison dynamics \cite{KF}.

Secondly, the spectrum can contract in the limit of strong resolvent
convergence (although it cannot expand) in contrast to norm resolvent
convergence, in which the spectrum can neither contract nor expand.
(See Theorem VIII.24(a) and the following discussion in \cite{RSi}.)
This property of strong resolvent convergence will be important in
the sequel; furthermore norm resolvent convergence is too strong for
our purposes: one can see that there are no analogues of some of our
results in this sense of convergence.

We also note that the results in \cite{Alb} concern sequences of
scaled Hamiltonians, whereas (due to our weaker notion of
convergence) we are not restricted in this way. In addition, the
scattering length formalism developed in Section 4 provides a
transparent physical interpretation for our convergence results. In
\cite{KF} these differences are discussed at greater length in the
case of short range (non-Coulombic) potentials.

We quickly review the classification of the self-adjoint extensions
of $\Hi$ along the lines of the discussion in \cite{Alb}. The
deficiency indices \cite{RSii} $n^\pm$ of $\Hi$ are defined to be the
dimensions of the deficiency subspaces $W^\pm = \ker (\Hia\mp i)$,
i.e. the number of independent $L^2$ solutions to
\begin{equation}
\Hia\psi=\pm i\psi.
\end{equation}
$\Hia$ may be  seen to act `classically' (cf. Proposition 2 in the
appendix to section X.1 of \cite{RSii}) i.e. such that any such
solution solves
\begin{equation}
\left(-\triangle + \frac{\gamma}{r}\right) \psi = \pm i\psi
\end{equation}
as a differential equation, and so usual techniques of partial
differential equations suffice. In the familar way, we use
the decomposition
\begin{equation}
L^2(\RR^3,d^3\rvec) = \bigoplus_{\ell=0}^\infty
L^2(\RR^+,r^2dr)\otimes {\cal K}_\ell    \label{eq:dcmp}
\end{equation}
where ${\cal K}_\ell$ is the subspace of $L^2({\Bbb S}^2,d\Omega)$
spanned by $Y_{\ell,-\ell},\ldots,Y_{\ell,\ell}$, to separate over
the basis of spherical harmonics. We make the unitary
transformation $U:L^2(\RR^+,r^2dr)\rightarrow L^2(\RR^+,dr)$
given by $(U\chi)(r)=r\chi(r)$ to obtain the radial
Hamiltonians for each angular momentum sector
\begin{equation}
h_\ell = -\frac{d^2}{dr^2} + \frac{\ell (\ell+1)}{r^2} +
\frac{\gamma}{r}
\label{eq:Hrad}
\end{equation}
which we define on $C_0^\infty(0,\infty)\subset L^2(\RR^+,dr)$.
(Studying the $h_\ell$ on this domain is actually equivalent to
studying $\Hi$ on $\tilde{\DD}$, the set of finite linear
combinations of terms of form $f(r)Y_{\ell,m}(\theta,\varphi)$ where
$f(r)\in C_0^\infty(0,\infty)$. However, this can be seen to yield
the same deficiency subspaces and self-adjoint extensions as are
obtained with $\DD =\Coinf$.)

Solving $h_\ell^*u_\ell^\pm=\pm iu_\ell^\pm$ as an ODE yields, as
solutions square integrable at infinity
\begin{equation}
u_\ell^\pm=
{\cal W}_{-i\gamma/2(\pm i)^{1/2};\ell+1/2}(-2i(\pm i)^{1/2}r)
\label{eq:ul}
\end{equation}
where ${\cal W}_{\kappa;\mu}(z)$ is a Whittaker function \cite{Grad}.
For $\ell\ge 1$, these functions are not square integrable at the
origin and so the $h_\ell$ are essentially self-adjoint. In the case
$\ell=0$, however, the functions $u_0^\pm$ are square integrable, so
$h_0$ has deficiency indices $\langle 1,1\rangle$ and a 1-parameter
family of self-adjoint extensions labelled by $U(1)$, or
equivalently (under an obvious correspondence) by the extended real
line $\RR\cup\{\infty\}$. These may be denoted\footnote{Our
parametrisation differs from that in \cite{Alb}. This will be
convenient later.} $h^L_0$ where  $L\in\RR\cup\{\infty\}$ and $h^L_0$
has domain
\begin{eqnarray}
D(h^L_0) & = &
\{u\in L^2(\RR^+)\mid u,u^\prime\in AC_{\rm loc}(\RR^+);
{}~u^{\prime\prime} + \gamma r^{-1}u\in L^2(\RR^+), \nonumber\\
& & \qquad\qquad\qquad
[L^{-1}+\gamma(\Psi(1)+\Psi(2))]u_0+u_1 = 0 \} \label{eq:dom}
\end{eqnarray}
where $u_0=\lim_{r\rightarrow 0^+} u(r)$,
$u_1=\lim_{r\rightarrow 0^+} r^{-1}(u(r)-u_0(1+\gamma
r\log(|\gamma|r)))$. Here $AC_{\rm loc}(\RR^+)$ denotes the set of
locally absolutely continuous functions \cite{RSi} on the positive
half-line, and $\Psi(z)\equiv d/dz \log\Gamma(z)$ is the di-gamma
function \cite{Grad}. $u_1$ can be regarded as a `Coulomb
modified' first derivative (from the right) of $u(r)$ at $r=0$.
Note that the usual Hamiltonian, with regular boundary conditions,
arises from the choice $L=0$.

The spectrum of $h^L_0$ in the case $\gamma<0$ was first discussed by
Rellich \cite{Rell}  (see \cite{Alb} for the $\gamma>0$ case). For
each value of $L$, there are infinitely many non-degenerate negative
eigenvalues of $h^L_0$. The eigenfunctions are precisely those
solutions $v(r)$ to
\begin{equation}
\left(-\frac{d^2}{dr^2} +\frac{\gamma}{r}\right)v = -\kappa^2v
\end{equation}
which are square integrable and lie in the domain~(\ref{eq:dom})
above. The first requirement entails that $v(r)$ is given (up to
normalisation) in terms of the Whittaker function by
\begin{equation}
v(r) = {\cal W}_{|\gamma|/2\kappa;\frac{1}{2}}(2\kappa r).
\end{equation}
Defining $v_0=\lim_{r\rightarrow 0^+} v(r)$, $v_1=\lim_{r\rightarrow
0^+} r^{-1}(v(r)- v_0(1+\gamma r\log(|\gamma|r)))$, we find that
\begin{equation}
\frac{v_1}{v_0} =\gamma\left[ \log\frac{2\kappa}{|\gamma|} - \Psi(1)
-\Psi(2)+\Psi(1+\gamma/2\kappa)-\kappa/\gamma\right] \label{eq:vl0}
\end{equation}
and hence (by our second requirement) that eigenvalues occur for
$E=-\kappa^2$ satisfying
\begin{equation}
\frac{a_o}{2L} = -\frac{1}{\gamma L} =
G\left(\frac{\gamma}{2(-E)^{1/2}}\right)
\label{eq:Leval}
\end{equation}
where the function $G$ is defined by
\begin{equation}
G(z)= \Psi(1+z)- \log |z| - 1/(2z) .
\end{equation}
$G(z)$ is everywhere increasing in $z\in\RR$ except at its poles,
located at $z=-n$ for $n=0,1,2,\ldots$, and so $G(z)$ increases from
$-\infty$ to $\infty$ between each consecutive pair of poles (see
Fig. 2.1). There is therefore precisely one zero $z_n$ of $G(z)$
between each consecutive pair of negative integers $z_n\in
(-n,-n+1)$ for $n=1,2,\ldots$ and precisely one solution
of~(\ref{eq:Leval}) for $\gamma/(2(-E)^{1/2})$ in the interval
$(z_{n+1},z_n]$ for each $n=1,2,\ldots$. We label the corresponding
energy by $E_n$. For $L>0$, there is also an eigenvalue $E_0$, with
$\gamma/(2(-E_0)^{1/2}\in (z_1,0)$.
In addition, the special
case $L=0$ -- which corresponds by~(\ref{eq:dom}) to the case of
regular boundary conditions at the origin -- gives rise to the usual
Bohr levels
\begin{equation}
E_{n}= -\frac{\gamma^2}{4n^2}\quad n=1,2,\ldots  .
\end{equation}
However, we see that by changing $L$, we can radically alter the
eigenvalues.  The eigenvalues $E_n$ for $n\ge 1$ are subject to
maximum fractional changes of order $1/n$, and for all $L>0$, there
is an additional eigenvalue $E_0$ lying well below the usual Bohr
levels. Thus the ground state and first few excited energy levels can
be quite sensitive to the  value of $L$.
However, if $|L|\ll a_0$, we can see (comparing with Fig. 2.1) that
the energy levels are little changed from the Bohr levels. Indeed, in
such cases, because $\Psi(1+\xi)=-(\xi+n)^{-1}+O(1)$ as
$\xi\rightarrow -n-1$ for all $n=1,2,\ldots$, we may write the second
equality  in~(\ref{eq:Leval}) in the approximate form
\begin{equation}
-\frac{1}{\gamma L}\approx -\frac{1}{\gamma/2(-E)^{1/2}+n}
\end{equation}
obtaining the approximate solutions
\begin{equation}
E_n=-\frac{\gamma^2}{4(n-\gamma L)^2} =
-\frac{\gamma^2}{4(n+2L/a_o)^2} .
\label{eq:apeval}
\end{equation}
Thus $2L/a_o$ plays the r\^{o}le of the Rydberg correction (or
quantum defect) of atomic physics \cite{Land}. In addition for $0<L\ll
a_0$, because $G(\xi)=-(2\xi)^{-1}+O(\log|\xi|)$ as $\xi\rightarrow
0$, we have an energy level $E_0$ lying well below the Bohr spectrum
\begin{equation}
E_0\approx -\frac{1}{L^2} .
\end{equation}
The presence of such a deeply bound state is initially puzzling, and
we shall return to this in Section 4.

As noted above, for $\ell\ge 1$, the operators $h_\ell$ are
essentially  self-adjoint on the domain $C_0^\infty(0,\infty)$ and so
there is a unique self-adjoint extension $\bar{h}_{\ell}$ whose
eigenvalues are simply the familiar Bohr levels:
\begin{equation}
E_{\ell,n} = -\frac{\gamma^2}{4(n+\ell)^2} \quad n=1,2,\ldots .
\end{equation}
By analogy, we will abuse notation and write $\bar{h}_0$ to denote
$h^0_0$. We may therefore assemble the radial Hamiltonians to
give the full Hamiltonian $H^L$ (recalling the
decomposition~(\ref{eq:dcmp})) defined by
\begin{equation}
H^L = U^*h^L_0 U\otimes {\Bbb I} \oplus
\bigoplus_{\ell=1}^\infty U^*\bar{h}_\ell U\otimes{\Bbb I}
\end{equation}
which acts on the $S$-wave as $h^L_0$ and as  $\bar{h}_{\ell}$
on the $\ell$'th sector for $\ell\ge 1$. The spectral properties of
$H^L$ are immediately given by the discussion above.

The case $H^0$ corresponds to the usual choice of
idealisation with regular boundary conditions. This operator is the
closure of $\Hi$ on $C_0^\infty(\RR^3)$ and would therefore result
if we treated the idealised problem in the same way as true
problems by assuming finiteness of the wavefunction as an axiom. We
emphasise that Theorem~\ref{Th:seq} shows that it would be
inconsistent to do this.

To summarise, there is a 1-parameter family of idealised Hamiltonians
arising as self-adjoint strong resolvent limits of sequences of
true problem Hamiltonians whose nuclear radii shrink to zero. The
idealisations $H^L$ are the self-adjoint extensions of
$-\triangle+\gamma/r$ on $\Coinf$ which are labelled by a single real
parameter $L\in \RR\cup\{\infty\}$ (later, we will identify
$L$ as a Coulomb modified scattering length). The idealisations
differ from the usual Hamiltonian $H^0$ only in the
$S$-wave, in which the low-lying energy levels can exhibit
significant fractional changes from the Bohr levels, depending on the
value of $L$. In addition, for all non-zero values of $L$,
there is an additional bound state lying below the usual Bohr
spectrum. We will discuss this bound state later in Section 4.

\sect{The Regular Boundary Condition}

We now turn to our rigorous justification of the regular boundary
condition. Central to this is the mild physical requirement that the
correction potential $V$ is positive and has nuclear radius $a\ll
a_0$. Under these assumptions, we will see that the spectrum of the
true problem is well approximated by the Bohr levels. In the
following, for any self-adjoint  operator $H$ bounded from below,
$\mu_n(H)$ denotes (when it exists), the $n$'th discrete eigenvalue
of $H$, counting with multiplicity and in increasing order.

Given positive $V\in\Cl$, we define the $S$-wave radial
true Hamiltonian $h_{0,\rm true}=\bar{h}_0+V$.
Suppose $V$ has nuclear radius $a$. Because $V$ is positive, the
$n$'th eigenvalue $\mu_n(h_{0,\rm true})$ is bounded below by the
$n$'th Bohr level $\mu_n(\bar{h}_0)$. Moreover, it is resonable to
suppose that an upper bound should be provided by the $n$'th
eigenvalue of $h^\prime_0$, the Hamiltonian describing an
impenetrable nucleus of radius $a$ which corresponds to the
heuristic potential
 \begin{equation}
V(r)=\left\{\begin{array}{cl} \infty & r<a \\ 0 & r\ge a.
\end{array}\right.
\end{equation}
In fact, this turns out to be the case. We define  $h^\prime_0$
defined  on $L^2([a,\infty),dr)$ by
\begin{equation}
h^\prime_0 = \overline{  \left(
-\frac{d^2}{dr^2} +\frac{\gamma}{r}\right) \upharpoonright\DD^\prime}
\label{eq:hp}
\end{equation}
where
\begin{equation}
\DD^\prime =\{\psi\mid (r-a)^{-1}\psi\in C_0^\infty([a,\infty))\}
\subset L^2([a,\infty),dr).
\end{equation}
The domain $\DD^\prime$ is chosen to ensure that all eigenfunctions
vanish at the radius $r=a$. In the Appendix, we prove that this
Hamiltonian has precisely one eigenvalue between each consecutive
pair of Bohr levels and no other eigenvalues. (In addition there is
essential spectrum $[0,\infty)$.) Note that this Hamiltonian is
defined on a different Hilbert space from $h_{0,\rm true}$. In
Section 6, we prove
\begin{Thm} \label{Th:posbd}
Let $h_{0,\rm true}=\bar{h}_0+V$ for some positive $V\in\Cl$ with
nuclear radius $a>0$ and define $h^\prime_0$ by~(\ref{eq:hp}) for this
choice of $a$. Then $\bar{h}_0$, $h_{0,\rm true}$ and $h^\prime_0$ are
bounded from below and possess infinitely many eigenvalues
satisfying
\begin{equation}
\mu_n(\bar{h}_0)\le \mu_n(h_{0,\rm true})\le \mu_n(h^\prime_0)
\label{eq:ineq}
\end{equation}
for each $n=1,2,\ldots$.
\end{Thm}

In the case $a\ll a_0$, the energy levels of $h^\prime_0$
are estimated in the Appendix as
\begin{equation}
\mu_n(h^\prime_0) \approx -\frac{\gamma^2}{4(n+2a/a_0)^2}.
\end{equation}
Hence~(\ref{eq:ineq}) may be re-written
\begin{equation}
-\frac{\gamma^2}{4n^2} \le \mu_n(h_{0,\rm true}) \lesssim
-\frac{\gamma^2}{4n^2} \left[ 1- \frac{4a}{na_0}\right]
\label{eq:ineq2}
\end{equation}
and so, using $a/a_0\sim 10^{-5}$, the maximum fractional error for
the physical hydrogen atom is of order $10^{-5}$.

Hence we see that the energy levels of the true problem are close to
the Bohr levels, provided the correction to the Coulomb potential is
positive and that the nuclear radius is smaller than the Bohr
radius.  Note that the bound~(\ref{eq:ineq2}) is considerably weaker
than the usual perturbative estimate of the energy shift due to the
finite size and structure of the nucleus \cite{Bethe}, which
assumes the nucleus is a smoothed out region of positive charge and
works to first order in perturbation theory, obtaining energy shifts
of order $(a/a_0)^2$. However, the result above is non-perturbative
and covers a much greater range of nuclear models, including many for
which first order perturbation theory would break down. In the next
section, we will describe how a general true problem can be modelled
by the idealised problem with potentially irregular boundary
conditions, and how the usual perturbative result may be derived
within this formalism.

As a second justification of the regular boundary condition, we
consider sequences of Hamiltonians with positively corrected Coulomb
potentials. Recall that $\Seq$ denotes the class of sequences of true
problem Hamiltonians $H_n=-\triangle+\gamma/r+V_n(r)$ whose nuclear
radii tend to zero. Our space of general regularisation schemes is
therefore $\Seq$. In general, Theorem~\ref{Th:seq} -- which defines
our notion of idealisation -- shows that class of the possible
self-adjoint limits of such sequences in the strong resolvent sense
is precisely the class of self-adjoint extensions of $\Hi$ on
$\Coinf$. Thus, in general, the limit (when it exists) is {\em not}
the self-adjoint extension with  regular boundary conditions.
However, restricting to the subclass $\Seq^+\subset\Seq$ of sequences
$H_n=-\triangle+\gamma/r+V_n(r)$ in which the potentials $V_n$ are
positive, it turns out that all such sequences converge in
the strong resolvent sense.
\begin{Thm}  \label{Th:poscvg}
All sequences in $\Seq^+$ converge in the strong
resolvent sense to $H^0$, the self-adjoint extension of $\Hi$ with
regular boundary conditions.
\end{Thm}
Thus any regularisation scheme in $\Seq^+$ selects the regular
boundary condition, providing our second justification.

\sect{Scattering Length Formalism}

In this section, we consider how a general true problem may be
approximated by an appropriate choice of idealised problem. The
starting point for our discussion is the low energy expansion of
Coulomb modified scattering theory developed by Lambert \cite{Lamb}.
This will enable us to examine the behaviour of both true problems and
idealisations at energies small in comparison with the nuclear scale,
i.e. $E\ll a^{-2}$. This regime covers the atomic spectrum, whose
characteristic scale is the Bohr radius, 5 orders of magnitude larger
than the atomic radius for the physical hydrogen atom. The `best fit'
boundary condition can then be found by matching the low energy
behaviour of the idealisation to the true problem. The procedure
below is the direct generalisation of the treatment of the
non-Coulombic case in \cite{KF}.

In the $S$-wave, we therefore examine the regular solution $u(r)$
to
\begin{equation}
\left\{ -\frac{d^2}{dr^2} + \frac{\gamma}{r} + V(r) \right\} u = k^2u
\label{eq:CS}
\end{equation}
where $V(r)$ obeys $r^2V(r)\rightarrow 0$ as $r\rightarrow 0$ and is
compactly supported within some radius $a$ of the origin. These
conditions allow all potentials in our class $\Cl$.

The Coulomb modified phase shift $\delta_0(k)$ is defined by writing
the asymptotic form of the regular solution to~(\ref{eq:CS}) as
\begin{equation}
u(r)\sim\sin (kr-(\gamma/(2k))\log 2kr +
\sigma_0(k) + \delta_0(k))
\end{equation}
where $\sigma_0(k)$ is given by
\begin{equation}
e^{2i\sigma_0(k)}=\frac{\Gamma(1+i\gamma/2k)}{\Gamma(1-i\gamma/2k)}.
\end{equation}
Defining
\begin{equation}
Z^c(k) = \pi\gamma(e^{\pi\gamma/k}-1)^{-1}\cot\delta_0(k) +
 \frac{\gamma}{2}
\left[ \Psi\left(\frac{i\gamma}{2k}\right) +
\Psi\left(-\frac{i\gamma}{2k}\right) \right] -
 \gamma\log \frac{|\gamma |}{2k} ,
\end{equation}
we have the (Coulomb modified) low energy expansion (see e.g.
\cite{Lamb})
\begin{equation}
Z^c(k) = -\frac{1}{L} + \frac{1}{2}r_o k^2 + O(k^4)   \label{eq:lowc}
\end{equation}
where $L$ and $r_o$ are the Coulomb modified scattering length
and effective range respectively. This is the analogue of the
familiar low energy expansion of scattering theory for short range
forces \cite{Newt}.

It can be shown \cite{Alb} that the scattering theory of the
idealised Hamiltonian $H^L$ is described by the $S$-wave low energy
expansion
\begin{equation}
Z^c(k) = -\frac{1}{L}.
\end{equation}
(For angular momenta $\ell\ge 1$, there is, of course, no scattering.)
Thus the operator $H^L$ has a low energy expansion which is exact at
lowest order. We are also able to interpret the parameter $L$ as the
(Coulomb modified) scattering length of $H^L$. It is therefore clear
from this and equation~(\ref{eq:lowc}) that scattering theory of $H^L$
describes the leading order scattering theory of any true problem with
scattering length $L$ at low energies. Moreover, because the low
energy expansion may be analytically continued to discuss bound
states, $H^L$ should also well-approximate the bound states of any
true problem with scattering length $L$ at sufficiently low energies.
It is therefore possible to select the `best fit' idealisation to a
given true problem simply by matching the scattering lengths.

In order to compute the scattering length of a given true problem,
it suffices to consider the regular solution $u(r)$ to~(\ref{eq:CS})
at zero energy, i.e. $k^2=0$. Lambert showed \cite{Lamb} that the
scattering length is then given by
\begin{equation}
L =
\frac{\phi(a)}{\theta(a)} -\frac{1}{\theta(a)^2} \left.
\left(\frac{u^\prime(r)}{u(r)} -
\frac{\theta^\prime(r)}{\theta(r)}\right)^{-1}\right|_{r=a} .
\label{eq:fitc}
\end{equation}
Because this formula allows us to select the best fit idealisation to
the true problem, we refer to it as the {\em fitting formula}.
Here,
\begin{equation}
\phi(r) = \left(\frac{r}{|\gamma|}\right)^{1/2}
J_1(2(|\gamma|r)^{1/2})   \label{eq:phi}
\end{equation}
and
\begin{equation}
\theta(r) =  -\pi (|\gamma|r)^{1/2} N_{1}(2(|\gamma|r)^{1/2})
\label{eq:theta}
\end{equation}
(where $J_1(z)$ and $N_1(z)$ are the Bessel and Neumann functions of
order 1 \cite{Grad}) are the regular and irregular solutions
to~(\ref{eq:CS}) in the special case $V\equiv 0$, $k=0$. Note that
$\phi$ and $\theta$ have Wronskian $\phi^\prime\theta -
\theta^\prime\phi \equiv 1$. The effective range may also be
determined using $u$, $\phi$ and $\theta$ \cite{Lamb}.

It is, of course important to know the range of energies over which
this approximation holds good. In \cite{KF}, this is referred to as
the question of {\em believability}. As a simple necessary condition
-- derived by requiring that the second term in~(\ref{eq:lowc}) be
small compared with the first -- we say that the idealisation is
believable at energy $E$ only if
\begin{equation}
|E| \ll \frac{1}{|Lr_o|} \label{eq:beli}
\end{equation}
where $L$ and $r_o$ are the scattering length and effective range of
the true problem.

As we noted in Section 2, for $L\gtrsim a_0$, the $S$-wave energy
levels exhibit significant fractional changes from the Bohr levels.
{}From this it is to be expected that the low-lying energy levels of
an atom differ markedly from the Bohr levels if the nucleus has
scattering length of the order of, or greater than, the Bohr radius.
It is therefore important to know whether the scattering length can
take such values. In fact a simple argument shows that for any $L\in
\RR\cup\{\infty\}$ there exist potentials $V(r)$ of arbitrarily small
support with scattering length $L$: let $\rvec\mapsto \psi(|\rvec |)$
be any smooth spherically symmetric function on $\RR^3$ with
$\psi(r)$ bounded away from zero for $r<a$ and
$\psi(r)=r^{-1}(\phi(r)-L\theta(r))$ for r$\ge a$. Then the potential
$V(r)$ defined by
\begin{equation}
V(r) = \psi^{-1}\triangle \psi - \frac{\gamma}{r} \label{eq:ptdf}
\end{equation}
is supported within $r<a$ and the effective potential
$\gamma/r+V(r)$ is smooth on $\RR^3$. Moreover, by~(\ref{eq:ptdf})
and the first property, $V(r)$ has scattering length $L$. We have
therefore proved
\begin{Thm} \label{Th:4.1}
For any $L\in\RR\cup\{\infty\}$, and any $a>0$, there exists a
potential $V\in\Cl$ with nuclear radius less than $a$ and with
(Coulomb modified)  scattering length $L$.
\end{Thm}

Thus there exist true problems with `nuclei' of arbitrarily small
radius whose energy levels at sufficiently small energy (i.e. for
large enough principal quantum number) are well-approximated by the
energy levels of $H^L$, for any $L\in \RR\cup\{\infty\}$. One can
use `believability' to determine the range of energies for which
this approximation is valid. Provided the effective range (and all
higher parameters in the low energy expansion) is small, the
approximation could include the energy levels of low principal
quantum number. Concrete examples are given in \cite{FK}.

We now restrict to the case of positive corrections $V\in\Cl$ to the
Coulomb potential. We take the nuclear radius to be $a$ and define
$u(r)$ and $\phi(r)$ as above, noting that they
satisfy
\begin{equation}
\phi(r) u^{\prime\prime}(r) -
u(r)\phi^{\prime\prime}(r) = V(r)\phi(r)u(r) .
\end{equation}
The first zero of $\phi(r)$ away from the origin occurs at
approximately $1.8a_0$, and that of $\theta(r)$ at $a_1\sim 0.6a_0$.
Restricting to the case $a<a_1$ (which therefore includes the
physical case $a\sim 10^{-5}a_0$), we have $\phi\ge 0$ on $(0,a)$ and
we may write
\begin{equation}
\phi(r)^2\frac{d}{dr}\left(\frac{u(r)}{\phi(r)}\right) =
\phi(r)u^\prime(r) -u(r)\phi^\prime(r) =
\int_0^r V(r^\prime)
\phi(r^\prime)u(r^\prime) dr^\prime  .\label{eq:wron}
\end{equation}
Because $r^2V(r)\rightarrow 0$ as $r\rightarrow 0$ and $u, \phi$ both
obey regular boundary conditions at the origin, we may divide through
by $\phi^2$ to yield an integrable function, leading to the following
integral equation for $u(r)$
\begin{equation}
u(r) = \phi(r) + \phi(r)\int_0^r dr^\prime (\phi(r^\prime))^{-2}
\int_0^{r^\prime} dr^{\prime\prime}
V(r^{\prime\prime})\phi(r^{\prime\prime})u(r^{\prime\prime}) .
\end{equation}
Since $\phi(r)> 0$ on $(0,a)$, we conclude that $u(r)$
is also strictly positive on this interval, for by the second equality
in~(\ref{eq:wron}), $u$ could vanish only if $u^\prime$ is positive.
Therefore, because $u(r)$ is initially positive, it must remain so on
this interval ($V(r)$ is sufficiently regular that $u(r)$
remains bounded on $(0,a)$). Hence we obtain the relation
\begin{equation}
\frac{u^\prime(r)}{u(r)} - \frac{\phi^\prime(r)}{\phi(r)}
= \frac{\int_0^r V(r^\prime) \phi(r^\prime)u(r^\prime) dr^\prime}
{\phi(r)u(r)} > 0 .
\end{equation}
Inserting this in the fitting formula~(\ref{eq:fitc}) and using the
fact that $\phi$ and $\theta$ have unit Wronskian and are positive
on $(0,a)$, we find that the Coulomb modified scattering length for
such a potential satisfies  \begin{equation}
0 < L < \frac{\phi(a)}{\theta(a)}.
\end{equation}
The trivial case $V(r)\equiv 0$ yields vanishing scattering length.
We have thus proved
\begin{Thm} \label{Th:slbd}
Let $V\in\Cl$ be positive and supported within $[0,a)$, where
$a<a_1$. Then the scattering length of $V$ is bounded by
\begin{equation}
0 \le L < \frac{\phi(a)}{\theta(a)} = L_{\rm max}. \label{eq:bd}
\end{equation}
\end{Thm}
Furthermore, if $a<a_0/8$, then a reasonable approximation to the
upper bound is provided by $a$, the nuclear radius, so we can write
\begin{equation}
0\le L \lesssim a .
\end{equation}

We can substitute this inequality into the approximate
expression~(\ref{eq:apeval}) in order to re-derive
equation~(\ref{eq:ineq2}) as a useful consistency check (although
the derivation in this section truncates the low energy expansion at
first order and is therefore less rigorous than our previous
derivation). In fact, we show in the Appendix that the $n+1$'st
eigenvalue of $h^{L_{\rm max}}_0$ provides an upper bound on the
$n$'th eigenvalue of $h^\prime_0$ and hence (by
Theorem~\ref{Th:posbd}) on the $n$'th eigenvalue of $h_{0,\rm
true}$ for a positive correction to the Coulomb potential. Note
that, for $0<L\lesssim a$, the idealisation $H^L$ also exhibits a
energy level $E_0(L)\sim -L^{-2}$ lying well below the Bohr spectrum,
as we saw in Section 2. However, we can use believability to show
that this is not a believable feature of the true system, at least in
the case where the effective range is of the order of the nuclear
radius, when the range of believability is
\begin{equation}
|E|\ll \frac{1}{a^2}< \frac{1}{L^2}
\end{equation}
so the bound state $E_0\approx -1/L^2$ falls outside this range. The
presence of this bound state is therefore a result of the idealisation
process which does not in this case correspond to any feature of the
true problem. This is reassuring, because it is, of course, impossible
for a positive correction to the Coulomb potential to introduce a new
energy level {\em below} the Bohr levels. This provides a simple
example of the manner in which the components of the scattering
length formalism and believability work together.

\sect{Higher Angular Momenta}

The discussion in Sections 3 and 4 demonstrated two features of the
true problem in the $S$-wave. Theorem~\ref{Th:posbd} showed that for
{\em positive} corrections to the Coulomb field of small nuclear
radii, the true spectrum is well approximated by the Bohr levels. On
the other hand, Theorem~\ref{Th:4.1} showed that there exist true
problems with arbitrarily small nuclear radius whose spectra depart
significantly from the Bohr levels. In this section, we examine the
true problem for higher angular momenta and discover that not only
are the Bohr levels stable against positive perturbations, they are
also remarkably stable against general perturbations in a manner we
will specify below. This can be traced directly to the fact that
$C_0^\infty(0,\infty)$ is a core for the radial Hamiltonians
$\bar{h}_\ell$, in contrast to the $\ell=0$ case, where the
deficiency indices are $\langle 1,1\rangle$.

We firstly examine the case $V(r)\ge 0$. Here, we have the analogue of
Theorem~\ref{Th:posbd}: for $a>0$ we define $h^\prime_{\ell}$ on
$L^2([a,\infty),dr)$ by
\begin{equation}
h^\prime_\ell = \overline{ \left(
-\frac{d^2}{dr^2} +\frac{\ell(\ell+1)}{r^2}
+\frac{\gamma}{r}\right)\upharpoonright\DD^\prime} \label{eq:hilhp}
\end{equation}
where
\begin{equation}
\DD^\prime=\{\psi\mid (r-a)^{-1}\psi\in
C_0^\infty([0,\infty))\} \subset L^2([a,\infty),dr).
\end{equation}
We have
\begin{Thm} \label{Th:hilpbd}
For any $\ell$, let $h_{\ell,\rm true}=\bar{h}_\ell+V$ for some
positive $V\in\Cl$  with nuclear radius $a>0$. Define $h^\prime_\ell$
by~(\ref{eq:hilhp}) for this choice of $a$.Then $\bar{h}_\ell$,
$h_{\ell,\rm true}$ and $h^\prime_\ell$ are bounded from below and
possess infinitely many eigenvalues satisfying
\begin{equation}
\mu_n(\bar{h}_{\ell}) \le \mu_n(h_{\ell,\rm true})\le
\mu_n(h^\prime_\ell)
\end{equation}
for each $n=1,2,\ldots$.
\end{Thm}

In the Appendix, we estimate the energy levels of $h^\prime_\ell$ as
\begin{equation}
\mu_n(h^\prime_\ell)= -\frac{\gamma^2}{4(n+\ell+\xi_n)^2}
\end{equation}
for $n=1,2,\ldots$, where
\begin{equation}
\xi_n \approx \frac{(n+2\ell)!}{(2\ell+1)!(2\ell)!(n-1)!}
\left(\frac{2}{n+\ell}\right)^{2\ell+1}
\left(\frac{a}{a_0}\right)^{2\ell+1}.
\end{equation}
Thus for positive corrections of small nuclear radius, the spectrum is
well approximated by the Bohr levels.

So far we have only considered spherically symmetric potentials $V$.
We state (but do not prove) an easy generalisation of
Theorems~\ref{Th:posbd} and~\ref{Th:hilpbd} to the general case.
\begin{Thm} \label{Th:nonsphsy}
Let $H_{\rm true}=H^0+V(\rvec)$ where $H^0$ is the idealisation with
regular boundary conditions and $V$ is positive and compactly
supported within some radius $a$ of the origin and such that
$\rvec\mapsto \gamma |\rvec|^{-1} + V(\rvec)$ is smooth on $\RR^3$.
Defining the $h^\prime_\ell$ for this choice of $a$, let
\begin{equation}
H^\prime = \bigoplus_{\ell=0}^\infty U^* h^\prime_\ell U.
\end{equation}
Then $H^0$, $H_{\rm true}$ and $H^\prime$ are all bounded from below
and possess infinitely many eigenvalues satisfying
\begin{equation}
\mu_n(H^0)\le \mu_n(H_{\rm true})\le \mu_n(H^\prime)
\end{equation}
for each $n=1,2,\ldots$.
\end{Thm}

Turning to case of general spherically symmetric potentials $V\in\Cl$,
the situation is as follows: the Bohr levels are preserved to
arbitrary accuracy by all true problems of sufficiently small nuclear
radius. To understand this precisely, consider a {\em gedanken}
experiment in which one examines true problems in the laboratory,
using a spectrometer which resolves energies down to tolerance
$\delta E$. Clearly, only finitely many spectral lines would be
observed. In the $S$-wave, one can find true problems of arbitrarily
small nuclear radius for which the spectral lines depart significantly
from those corresponding to the Bohr levels. In the sectors $\ell\ge
1$, however, all true problems of sufficiently small support would
apparently yield the spectral lines of the Bohr levels (up to the
tolerance of the spectrometer) with the additional possibility of
extra spectral lines. The important point is that the structure of
the Bohr levels is substantially preserved, in contrast to the
$\ell=0$ case.

These statements follow from the following result, which we prove in
Section 6.
\begin{Thm} \label{Th:hilgen}
Suppose $\ell\ge 1$ and
let $\lambda$ be an element of the spectrum of $\bar{h}_\ell$. Given
$\epsilon>0$, there exists a radius $a_{\ell}(\lambda,\epsilon)>0$
such that, for any potential $V\in\Cl$ supported within
$[0,a_{\ell}(\lambda,\epsilon))$
\begin{equation}
{\rm dist} (\lambda, \sigma(h_{\ell,true})) <\epsilon
\end{equation}
where $h_{\ell,\rm true}=\bar{h}_\ell+V$ and ${\rm
dist}(x,Y)=\inf_{y\in Y}\|x-y\|$.
\end{Thm}

This result is a direct consequence of the fact that
$C_0^\infty(0,\infty)$ is a core for $\bar{h}_\ell$. Our statement
above is an immediate corollary of Theorem~\ref{Th:hilgen}:
\begin{Cor} \label{Co:cor}
Suppose $\ell\ge 1$. Given $\epsilon>0$, there exists
$a_\ell(\epsilon)>0$ such that, for all $V\in\Cl$ supported within
$[0,a_\ell(\epsilon))$,
\begin{equation}
{\rm dist} (\lambda, \sigma(h_{\ell,true})) <\epsilon \label{eq:6.4}
\end{equation}
for all $\lambda\in\sigma (\bar{h}_\ell)$, where $h_{\ell,\rm
true}=\bar{h}_\ell+V$.
\end{Cor}
{\em Proof:} Given $\epsilon>0$, equation~(\ref{eq:6.4}) is
non-trivial only for finitely many eigenvalues $\lambda$ of
$\bar{h}_\ell$. $a_\ell(\epsilon)$ may then be taken to be the
minimum value of the $a_\ell(\lambda,\epsilon)$ furnished by
Theorem~\ref{Th:hilgen} over these $\lambda$. $\square$

An important feature of the above results is that they do not
preclude the existence of extra energy levels, in addition to those
close to Bohr levels. (This arises because our proof in Section 6
makes use of strong resolvent convergence.) Indeed, one may see that
such levels can occur by considering a potential well within the
nucleus which is gradually deepened gradually pulling bound state
well below the Bohr spectrum. The remarkable feature of these
results is that even during such a process, the Bohr levels remain
well approximated (up to given accuracy) by eigenvalues of the true
problem in angular momentum sectors $\ell\ge 1$.

The principal limitation of Theorem~\ref{Th:hilgen} and
Corollary~\ref{Co:cor} is that they are pure
existence results; we have gained no information on the magnitude of
$a_\ell(\lambda,\epsilon)$. Numerical investigations, however, suggest
that, for example, all true problems with nuclear radius $a\sim
10^{-5}\times a_0$ have $P$-wave energy levels within 2\% of
each of the lowest five $P$-wave Bohr levels.

\sect{Rigorous Proofs} \label{sec:cvgnce}

In this section, we provide the proofs of the theorems stated above.
We begin by developing a sufficient condition for a sequence $A_n$ of
self-adjoint operators to converge to a self-adjoint limit $A$. Our
condition is only slightly weaker than that given by Theorem VIII.26
in \cite{RSi} (which we will also use) and the proof is similar. In
the sequel we will often use the concepts of the graph of an operator
and also the {\em strong graph limit} of a sequence of operators:
\begin{Def}
The {\em graph} $\Gamma(A)$ of an operator $A$ on a Hilbert space
$\HH$ is the set of pairs $\langle \psi,A\psi\rangle\in\HH\times\HH$
where $\psi$ runs through the domain of $A$. If $\{A_n\}$ is a
sequence of operators on $\HH$, the {\em strong graph limit} of the
$A_n$ is the set of pairs $\langle \phi,\psi\rangle\in\HH\times\HH$
for which there exists a sequence $\phi_n$ satisfying $\phi_n\in
D(A_n)$, $\phi_n\rightarrow\phi$ and $A_n\phi_n\rightarrow \psi$.
\end{Def}

We now give our sufficient condition for strong resolvent
convergence.
\begin{Prop} \label{Pr:suff}
Let $\{A_n\}$ and $A$ be self-adjoint on $\HH$ and let $\DD$ be a core
for $A$. If the graph of $A$ restricted to $\DD$ is contained in the
strong graph limit of the $A_n$, then $A_n\rightarrow A$ in the
strong resolvent sense.
\end{Prop}
{\em Proof:} Take $\phi\in\DD$. Then there exists a sequence
$\{\phi_n\}$, $\phi_n\in D(A_n)$ with $\phi_n\rightarrow \phi$ and
$A_n\phi_n\rightarrow A\phi$. For any $z\in\CC\backslash\RR$, we
have
\begin{eqnarray}
\left[ (A_n+z)^{-1} - (A+z)^{-1}\right](A+z)^{-1} & = &
(A_n+z)^{-1}\left[  (A+z)\phi-(A_n+z)\phi_n\right] \nonumber \\
& & +\phi_n-\phi \\
&  \longrightarrow & 0   \nonumber
\end{eqnarray}
as $n\rightarrow\infty$, because the $(A_n+z)^{-1}$ are a uniformly
bounded family of operators. Thus we have
\begin{equation}
(A_n+z)^{-1}\psi \rightarrow (A+z)^{-1}\psi
\end{equation}
for all $\psi\in {\rm Ran}
((A+z)\upharpoonright\DD)$, which is dense in $\HH$ because $\DD$
is a core for $A$. The result therefore extends to all $\psi\in\HH$
because the $(A_n+z)^{-1}$ and $(A+z)^{-1}$ are uniformly bounded.
$\square$

We will also need Theorem VIII.26 of \cite{RSi}:
\begin{Prop} \label{Pr:iff}
Let $\{A_n\}$ and $A$ be self-adjoint on $\HH$.
$A_n\rightarrow A$ in the strong resolvent sense if and only if the
graph of $A$ is the strong graph limit of the $A_n$.
\end{Prop}

We study two classes of sequences of true problem Hamiltonians
whose nuclear radii shrink to zero: $\Seq$ and $\Seq^+$. A
particularly useful feature of such sequences is that, for any smooth
wavefunction $\psi$ compactly supported away from the origin,
$H_n\psi\rightarrow (-\triangle+\gamma/r)\psi$ as
$n\rightarrow\infty$, because the support of the correction to the
Coulomb field eventually lies closer to the origin than the support
of $\psi$. In terms of graphs, this shows that the graph of
$\Hi=-\triangle+\gamma/r$ on $\Coinf$ is contained in the strong
graph limit of any sequence in $\Seq$.

In the results below, it will be useful to have available explicit
cores $\DD^L$ for the self-adjoint extensions $H^L$ of $\Hi$. To
this end, we define the {\em mollified zero energy wavefunction}
$\chi^L$ of $H^L$ by \begin{equation}
\chi^L(r) = g(r)r^{-1}\left(\phi(r)-L\theta(r)\right)
\end{equation}
where $g(r)$ is a smooth mollifier satisfying $g(r)=1$ for $r<a_0$ and
vanishing for $r>2a_0$. Clearly $\chi^L\in L^2(\RR^3,d^3\rvec)$. We
now define
\begin{equation}
\DD^L = C_0^\infty(\RR^3\backslash\{0\}) +
\{\lambda\chi^L\mid\lambda\in\CC\}
\end{equation}
and show that it is a core for $H^L$.
\begin{Lem} \label{Le:core}
$\DD^L$ is a core for $H^L$.
\end{Lem}
{\em Proof:} $H^L\upharpoonright\DD^L$ is clearly symmetric
and extends $\Hi$, so $\Hi^*$ extends $(H^L\upharpoonright\DD^L)^*$.
Therefore the deficiency subspaces of $H^L\upharpoonright\DD^L$ are
contained within those of $\Hi$. However, a simple integration by
parts argument shows that neither of the functions
$\psi_0^\pm=r^{-1}u_0^\pm$ are in the domain of
$(H^L\upharpoonright\DD^L)^*$, where $u_0^\pm$ are defined
in~(\ref{eq:ul}). Since $\psi_0^\pm$ span the two 1-dimensional
deficiency subspaces of $\Hi$, it is clear that the deficiency
indices of $H^L$ on $\DD^L$ are $\langle 0,0\rangle$, so that $\DD^L$
is a core for $H^L$. $\square$

We begin with Theorem~\ref{Th:seq} which demonstrates the connection
between the limit points of sequences in $\Seq$ and the self-adjoint
extensions of $\Hi$.
\begin{Recap}{\ref{Th:seq}}
Let $\Hi=-\triangle+\gamma/r$ on $\Coinf$.
(a) Let $\{H_n\}\in\Seq$. If $\{H_n\}$ has a self-adjoint
limit $H$ in the strong resolvent sense, then $H$ is a self-adjoint
extension of $\Hi$. (b) Furthermore, all self-adjoint extensions of
$\Hi$ arise as the strong resolvent limits of sequences in $\Seq$.
\end{Recap}
{\em Proof:} (a) Suppose that $\{H_n\}$ has a self-adjoint limit. Then
by Proposition~\ref{Pr:iff}, the strong graph limit of the $H_n$ is
equal to the graph of $H$. Furthermore. our observation above shows
that the graph of $H$ necessarily contains the graph of $\Hi$. Hence
$H$ is a self-adjoint extension of $\Hi$.

\noindent
(b) Now suppose that $H^L$ is the self-adjoint extension of $\Hi$ with
scattering length $L$. By Lemma~\ref{Le:core}, we know that $\DD^L$ is
a core for $H^L$. Proposition~\ref{Pr:suff} shows us that it it is
enough to construct a sequence $\{H_n\}\in\Seq$ whose strong graph
limit contains the graph of $H^L$ restricted to $\DD^L$. In fact, as
we know that the strong graph limit of any sequence in $\Seq$ contains
the graph of $\Hi$, it is enough to construct such a sequence whose
strong graph limit contains the pair $\langle\chi^L,
H^L\chi^L\rangle$. In order to do this, we employ similar arguments
to those used in the proof of Theorem~\ref{Th:4.1} to construct a
sequence of potentials of decreasing support all of which have
scattering length $L$. Specifically, let the $\chi_n(r)$ be smooth
spherically symmetric compactly supported functions on $\RR^3$
satisfying
\begin{enumerate}
\item $\chi_n(r)=\chi^L(r)$ for $r>a_0/n$
\item $\chi_n\rightarrow\chi^L$ in $L^2(\RR^3,d^3\rvec)$
\item $\chi_n(r)$ is bounded away from zero for $r<a_0/n$.
\end{enumerate}
We then define the potentials $V_n(r)$ by
\begin{equation}
V_n(r) =\left\{\begin{array}{cl} \chi_n^{-1}\triangle\chi_n -\gamma/r
& {\rm for}~ r<a_0 \\ 0 & {\rm for}~ r>a_0. \end{array} \right.
\end{equation}
It is easy to see that the $V_n\in\Cl$ and so $\{H_n\}$ given by
$H_n=-\triangle+\gamma/r+V_n(r)$ is an element of $\Seq$. Furthermore,
each term in the sequence has scattering length $L$. It is clear that
$H_n\chi_n=H^L\chi^L$ for all $n$ (for within radius $a_0$, both
sides vanish, whilst for $r>a_0$, $\chi_n$ and $\chi^L$ are equal,
and $H_n$ and $H^L$ have the same action). Moreover, because
$\chi_n\rightarrow \chi^L$, the strong graph limit of the $H_n$
contains $\langle \chi^L,H^L\chi^L\rangle$. By our remarks above,
this suffices to prove that $H_n\rightarrow H$ in the strong
resolvent sense. $\square$

Next, we consider the situation when we restrict from $\Seq$ to
$\Seq^+$ and prove Theorem~\ref{Th:poscvg}. (We defer the proof of
Theorem~\ref{Th:posbd} to the end of this section.)
\begin{Recap}{\ref{Th:poscvg}}
All sequences in $\Seq^+$ converge in the strong resolvent sense to
$H^0$, the self-adjoint extension of $\Hi$ with regular boundary
conditions.
\end{Recap}
{\em Proof:} Let $H_n$ be a sequence in $\Seq^+$. In exactly the same
way as in the proof of Theorem~\ref{Th:seq}(b), to show that
$H_n\rightarrow H^0$, it is enough to show that $\langle
\chi^0,H^0\chi^0\rangle$ is contained in the strong graph limit of
the $H_n$, where $\chi^0$ is the mollified zero energy wavefunction
of $H^0$. To prove this, let $\psi_n$ be the generalised zero energy
wavefunctions for the $H_n$, normalised such that
$\psi_n(r)=r^{-1}(\phi(r)-L_n\theta(r))$ for $r>a_n$, where $a_n$ is
the radius of support and $L_n$ the scattering length of $V_n$. Then
$\chi_n(r)=g(r)\psi_n$ is in $L^2(\RR^3,d^3\rvec)$.

Moreover, $\chi_n\in D(H_n)$ for all $n$ and
\begin{equation}
H^0\chi^0-H_n\chi_n= L_n r^{-1}\left(
g^{\prime\prime}\theta+2g^\prime\theta^\prime\right)\vartheta(r-a_n)
\end{equation}
where $\vartheta(x)$ is the Heaviside function. Thus, because the
term in parentheses is smooth and compactly supported, this tends to
zero, since, by Theorem~\ref{Th:slbd}, $0\le
L_n<\phi(a_n)/\theta(a_n)\rightarrow 0$. Thus $H_n\chi_n\rightarrow
H^0\chi^0$.

It therefore remains to show that $\chi_n\rightarrow\chi^0$. We know
that $\chi^0-\chi_n=L_ng(r)r^{-1}\theta(r)$ for $r>a_n$ and that
$g(r)r^{-1}\theta(r)$ has finite norm. Hence
\begin{equation}
\int_{a_n}^\infty |\chi^0(r)-\chi_n(r)|^2 r^2dr \rightarrow 0
\end{equation}
as $n\rightarrow \infty$. Furthermore, for positive potentials, we
know from the proof of Theorem~\ref{Th:slbd} that $r \chi_n(r)$ is
increasing in $r$  on $(0,a_n)$ for sufficiently small $a_n$, which
implies that $|r\chi_n(r)|$ is bounded above by
$\phi(a_n)-L_n\theta(a_n)$ on this interval and hence that
$\int_0^{a_n} |\chi_n-\chi^0|^2r^2dr\le \int_0^{a_n} (|r\chi(r)|^2
+|r\chi_n(r)|^2)dr \rightarrow 0$.

We therefore have $\chi_n\rightarrow \chi^0$, which entails that
$\langle \chi^0,H^0\chi^0\rangle$ is in the strong graph limit of the
$H_n$, yielding the required result. $\square$

As we noted in Section 2, whilst motivating the choice of strong
resolvent convergence, the spectrum cannot expand in the limit of
strong resolvent convergence. The precise statement of this result is:
\begin{Prop}
{\em (Theorem VIII.24(a) in \cite{RSi})}  Let $\{A_n\}$
and $A$ be self-adjoint and suppose $A_n\rightarrow A$ in the strong
resolvent sense. Then if $\lambda\in\sigma(A)$, there exist
$\lambda_n\in \sigma(A_n)$ such that $\lambda_n\rightarrow \lambda$.
\label{Pr:nexp}
\end{Prop}

This proposition allows us to prove Theorem~\ref{Th:hilgen}.
\begin{Recap}{\ref{Th:hilgen}}
Suppose $\ell\ge 1$ and
let $\lambda$ be an element of the spectrum of $\bar{h}_\ell$. Given
$\epsilon>0$, there exists a radius $a_{\ell}(\lambda,\epsilon)>0$
such that, for any potential $V\in\Cl$ supported within
$[0,a_{\ell}(\lambda,\epsilon))$
\begin{equation}
{\rm dist} (\lambda, \sigma(h_{\ell,true})) <\epsilon
\end{equation}
where $h_{\ell,\rm true}=\bar{h}_\ell+V$ and ${\rm
dist}(x,Y)=\inf_{y\in Y}\|x-y\|$.
\end{Recap}
{\em Proof:} Suppose otherwise. Then one could construct a sequence
$V_n(r)$ of potentials in $\Cl$ with $V_n$ compactly supported within
$[0,a_0/n]$, and hence a sequence of operators
$h_{\ell,n}=\bar{h}_\ell+V_n(r)$ such that
${\rm dist}(\lambda,\sigma(h_{\ell,n})) \ge\epsilon$ for all $n$. But
it is clear that the strong graph limit of this sequence contains the
graph of $\bar{h}_\ell$ restricted to $C_0^\infty(0,\infty)$, on
which it is essentially self-adjoint. By Proposition~\ref{Pr:suff},
therefore, we have $h_{\ell,n}\rightarrow \bar{h}_{\ell}$.
Proposition~\ref{Pr:nexp} shows that there must therefore exist a
sequence $\{\lambda_n\}$, with $\lambda_n\in\sigma(h_{\ell,n})$ and
$\lambda_n\rightarrow \lambda$. But this contradicts the fact that
${\rm dist}(\lambda,\sigma(h_{\ell,n})) \ge\epsilon$ for all $n$ by
construction. $\square$

We conclude with the proof of Theorem~\ref{Th:hilpbd}, which includes
Theorem~\ref{Th:posbd} as a special case. This requires some more
definitions. We write $Q(A)$ for the quadratic form domain of an
operator $A$ and employ the usual abuse of notation by writing
$\inner{\phi}{A\psi}$ for the value of the quadratic form associated
to $A$ acting on $\phi,\psi$. By $\mu_n(A)$, we denote (if it exists)
the $n$'th eigenvalue of a self-adjoint operator $A$ which is bounded
from below, counting in increasing order and with multiplicity.

We shall use the min-max principle (Theorem XIII.2 in
\cite{RSiv}) to determine the eigenvalues of the operators at hand.
Specifically, for $A$ a self-adjoint operator bounded from below,
define \begin{equation}
\nu_n(A) =
\sup_{\varphi_1,\ldots,\varphi_{n-1}}
\inf_{\psi\in U_A(\varphi_1,\ldots,\varphi_{n-1})}
\inner{\psi}{A\psi}
\end{equation}
with
\begin{equation}
U_A(\varphi_1,\ldots,\varphi_m)  = \{\psi\mid \|\psi\|=1, \psi\in
Q(A), \psi\in [\varphi_1,\ldots,\varphi_m]^\perp\}
\end{equation}
where $[\varphi_1,\ldots,\varphi_m]$ denotes the linear span of
$\varphi_1,\ldots,\varphi_m$. The min-max principle states that, for
each $n$, {\em either} there are at least $n$ eigenvalues
$\mu_1(A),\ldots,\mu_n(A)$ of $A$ given by $\mu_r(A)=\nu_r(A)$ for
$r=1,\ldots,n$ below the essential spectrum of $A$, {\em or}
$\nu_n(A)=\inf\sigma_{\rm ess}(A)$ in which case there are at most
$n-1$ eigenvalues below the essential spectrum given by
$\mu_r(A)=\nu_r(A)$ for $r=1,\ldots,n-1$ and
$\nu_m(A)=\inf\sigma_{\rm ess}(A)$ for all $m\ge n$. Thus if $A$ has
infinite discrete spectrum below the essential spectrum (which holds
for the operators we consider) these eigenvalues are given directly
by the min-max principle. In order to compare the eigenvalues of
operators on different Hilbert spaces (as required for
Theorems~\ref{Th:posbd} and~\ref{Th:hilpbd}) we make the following
definition, which is a mild extension of a definition in Section
XIII.2 of \cite{RSiv}.
\begin{Def} \label{De:le}
Let $A$ and $B$ be
self-adjoint and bounded below on Hilbert spaces $\HH_1,\HH_2$
respectively, where $\HH_2$ is isometrically embedded in $\HH_1$ by
$\Emb$. We write $A\le B$ if and only if $\Emb Q(B)\subseteq Q(A)$ and
$\inner{\Emb\psi}{A\Emb\psi}_{\HH_1}\le\inner{\psi}{B\psi}_{\HH_2}$
for all $\psi\in Q(B)$.
\end{Def}

As an immediate consequence of this definition and the min-max
principle, we have (cf. Section XIII.2 in \cite{RSiv})
\begin{Prop} \label{Pr:le}
Let $A$ and $B$ be self-adjoint and bounded below on on Hilbert spaces
$\HH_1,\HH_2$ respectively, where $\HH_2$ is isometrically embedded
in $\HH_1$ by $\Emb$. If $A\le B$ then $\nu_n(A)\le\nu_n(B)$ for
each $n$.
\end{Prop}

Theorems~\ref{Th:posbd} and~\ref{Th:hilpbd} turn on the following
lemma. Recall that $\bar{h}_0$ denotes $h^0_0$, the $S$-wave
radial Hamiltonian with regular boundary conditions, which is
bounded below.
\begin{Lem} \label{Le:ineq}
Let $h_{\ell,\rm true}=\bar{h}_\ell+V$ for some positive
$V(r)\in\Cl$, supported within $[0,a)$, for $a>0$. For each
$\ell=0,1,2,\ldots$, the operators $\bar{h}_\ell,  h_{\ell,\rm true}$,
and $h^\prime_\ell$ are bounded from below and satisfy
$\bar{h}_\ell\le h_{\ell,\rm true} \le h^\prime_\ell$.
\end{Lem}
{\em Proof:} By hypothesis on the class of potentials $\Cl$ from which
$V$ is taken, $Q(\bar{h}_\ell) = Q(h_{\ell,\rm true})$, and since
$V(r)\ge 0$, we have $h_{\ell, \rm true}$ bounded
below and $\bar{h}_\ell\le h_{\ell,\rm true}$ (we are in the special
case of Definition~\ref{De:le} in which $\HH_1=\HH_2$ and $\Emb$ is
the identity). To prove the second inequality, let $\Emb$ be the
isometry $\Emb :L^2([a,\infty),dr)\mapsto L^2(\RR^+,dr)$ given by
\begin{equation}
(\Emb\psi)(r) =
\left\{\begin{array}{cl}\psi(r)  & {\rm for}~r\ge a \\
                        0 & {\rm for}~r<a. \end{array} \right.
\end{equation}
$Q(h^\prime_\ell)$ is the set of $\psi$ such that
\begin{equation}
\int_a^\infty \left\{
\left|\frac{d}{dr}\psi\right|^2  +\left(
\frac{\ell(\ell+1)}{r^2} +\frac{\gamma}{r}\right) |\psi |^2
\right\}dr<\infty
\end{equation}
and it is clear that any such $\psi$ satisfies
\begin{equation}
\int_0^\infty \left\{
\left|\frac{d}{dr}\Emb\psi\right|^2  +\left(
\frac{\ell(\ell+1)}{r^2} +\frac{\gamma}{r}\right) |\Emb\psi |^2
\right\} dr<\infty
\end{equation}
as $\Emb\psi$ is continuous and piecewise once continuously
differentiable on $\RR^+$. Thus $\Emb Q(h^\prime_\ell)\subset
Q(\bar{h}_\ell) = Q(h_{\ell,\rm true})$. Clearly we have
$\inner{\Emb\psi}{h_{\ell,\rm true}\Emb\psi} =
\inner{\psi}{h^\prime_\ell\psi}$ for all $\psi\in
Q(h^\prime_\ell)$, so $h^\prime_{\ell}$ is bounded below and
$h_{\ell,\rm true} \le h^\prime_\ell$. $\square$

We now prove Theorem~\ref{Th:hilpbd}.

\begin{Recap}{\ref{Th:hilpbd}}
For any $\ell$, let $h_{\ell,\rm true}=\bar{h}_\ell+V$ for some
positive $V\in\Cl$  with nuclear radius $a>0$. Define $h^\prime_\ell$
by~(\ref{eq:hilhp}) for this choice of $a$.Then $\bar{h}_\ell$,
$h_{\ell,\rm true}$ and $h^\prime_\ell$ are bounded from below and
possess infinitely many eigenvalues satisfying
\begin{equation}
\mu_r(\bar{h}_{\ell}) \le \mu_r(h_{\ell,\rm true})\le
\mu_r(h^\prime_\ell)
\end{equation}
for each $n=1,2,\ldots$.
\end{Recap}
{\em Proof:} From Lemma~\ref{Le:ineq} and Proposition~\ref{Pr:le},
we have $\nu_n(\bar{h}_\ell)\le\nu_n(h_{\ell,\rm
true})\le\nu_n(h^\prime_\ell)$. $\bar{h}_\ell$ has infinite discrete
spectrum below its essential spectrum and in the Appendix, we show
that the same is true for $h^\prime_\ell$, so the min-max principle
gives $\mu_n(\bar{h}_\ell)=\nu_n(\bar{h}_\ell)$ and
$\mu_n(h^\prime_\ell)=\nu_n(h^\prime_\ell)$. A
further application of the min-max principle shows that $h_{\ell,\rm
true}$ also has infinite discrete spectrum below its essential
spectrum, thus yielding the required inequalities. $\square$

\sect{Conclusion}

We first briefly consider the relation of the present paper to our
work with Kay on the large scale effects of large objects in quantum
mechanics \cite{KF}. More discussion is given in \cite{FK}. The
`principle of sensitivity' states that the fine details of a small
object in the true problem (here the specfic nuclear model chosen)
can only change the large scale behaviour of the system if a family of
idealised problems with differing large scale behaviour can be found.
In the case in hand, we have seen in Section 2 that there is such a
family of idealised problems in the $S$-wave, whilst for angular
momenta $\ell\ge 1$, there is a unique idealised problem in each
sector. Furthermore, Theorem~\ref{Th:4.1} shows that true problems of
arbitrarily nuclear radius can have spectra differing markedly from
the Bohr levels in the $S$-wave, so there is good agreement with the
principle of sensitivity in this sector. The situation for $\ell\ge
1$ is more subtle. We have seen (Corollary~\ref{Co:cor} to
Theorem~\ref{Th:hilgen}) that all true problems of sufficiently small
nuclear radius preserve the Bohr levels to arbitrary accuracy;
however, additional energy levels (termed `rogue eigenvalues' in
\cite{KF}) may also be introduced. There is, therefore, a sense in
which the spectrum of the idealised problem is highly stable against
general perturbations in these sectors, in accordance with the
principle of sensitivity. Our considerations have thus demonstrated
broad agreement with, and have clarified the scope of, the principle
of sensitivity.

Our results in Section 5 also lead to an interesting viewpoint on the
phenomenon of `accidental degeneracy' \cite{Fock,Barg,BItz,Kaln};
i.e. the fact that the Bohr levels for $\ell\ge 1$ are coincident
with energy levels in the $S$-wave. This phenomenon is due to the
conservation of the Laplace-Runge-Lenz vector, a symmetry of the
idealised Hamiltonian with regular boundary conditions which is not
shared by any true problem. If the nucleus is modelled by a positive
correction to the Coulomb potential of small nuclear radius,
Theorems~\ref{Th:posbd} and~\ref{Th:hilpbd} demonstrate that the
degeneracy is approximately maintained. More generally one might
expect that the degeneracy would be badly broken. Whilst the $S$-wave
energy levels can indeed depart significantly from the Bohr levels,
Theorem~\ref{Th:hilgen} provides the surprising result that
degeneracy is maintained to arbitrary accuracy for the eigenvalues in
$\ell\ge 1$ for all true problems of sufficiently small nuclear
radius, modulo the presence of rogue eigenvalues. For angular
momentum higher than zero, accidental degneracy is no accident!

Our justification of the regular boundary condition was premised on
the assumptions that the nuclear structure is spherically symmetric
and can be modelled by a positive correction of compact support to the
Coulomb potential. Theorem~\ref{Th:nonsphsy} indicates how the
requirement of spherical symmetry may be relaxed. We also note that
many non-everywhere-positive potentials have scattering lengths
$L\lesssim a$ (for instance, if the potential is `only slightly'
negative) and that one could therefore use the scattering length
formalism of Section 4 to estimate the energy levels as being close
to the Bohr levels. However to do this would necessitate a careful
study of `believability'. It would be interesting to make rigorous
statements about such cases and also to relax the requirement of
compact support to allow, for example, exponentially decaying tails
outside the nuclear radius.

To summarise: by identifying the different r\^oles of true and
idealised problems in quantum mechanics, we have clarified the issue
of which boundary conditions should be imposed at the Coulomb
singularity of the idealised hydrogen atom. Under
mild physical restrictions on the true problem, we have provided a
rigorous justification for the regular boundary condition.
Furthermore, we have developed a formalism for matching general true
problems to (potentially) irregular boundary conditions, and have
demonstrated that the entire 1-parameter family of boundary are
required to model the full class of true problems, which (for
arbitrarily small nuclear radius) can have $S$-wave energy levels
differing markedly from the Bohr levels.

{\em Acknowledgements:} I am grateful to Dr B.S. Kay, my research
supervisor, for suggesting this problem to me and for suggesting the
application of the scattering length formalism. I also thank him for
many useful conversations. I thank the Institute for Theoretical
Physics, University of Berne, Switzerland and also the Department
of Mathematics, University of York, U.K. for hospitality whilst
part of this work was carried out. I thank Churchill College,
Cambridge, for financial support under a Gateway Studentship.

\appendix
\renewcommand{\thesection}{Appendix:}
\sect{Spectral Properties of $h^\prime_\ell$}
\renewcommand{\thesection}{A}

In this appendix we prove the following statements concerning
the spectrum of $h^\prime_\ell$ on $L^2([a,\infty),dr)$ (recall that
$\mu_n(H)$ denotes the $n$'th eigenvalue of $H$, counting in
increasing order and with multiplicity, where $H$ is self-adjoint and
bounded below):
\begin{enumerate}
\item $h^\prime_\ell$ has infinite discrete spectrum below its
essential spectrum  $[0,\infty)$. Moreover, $\mu_n(h^\prime_\ell)\in
[\mu_n(\bar{h}_\ell),\mu_{n+1}(\bar{h}_\ell))$.
\item In $\ell=0$, $\mu_n(h^\prime_0)\le\mu_{n+1}(h^{L_{\rm max}}_0)$
where $L_{\rm max} = \phi(a)/\theta(a)$, provided $a<a_1$, where $a_1$
is the first non-trivial zero of $\theta(r)$ away from the origin.
\item For $a\ll a_0$,
\begin{equation}
\mu_n(h^\prime_\ell) = -\frac{\gamma^2}{4(n+\ell+\xi_n)^2}
\end{equation}
for
\begin{equation}
\xi_n \approx \frac{(n+2\ell)!}{(2\ell+1)!(2\ell)!(n-1)!}
\left(\frac{2}{n+\ell}\right)^{2\ell+1}
\left(\frac{a}{a_0}\right)^{2\ell+1}.
\end{equation}
\end{enumerate}

To prove these results, we first write $u_\kappa(r)$ for the
solution to
\begin{equation}
-u^{\prime\prime} + \left(\frac{\ell(\ell+1)}{r^2}
+\frac{\gamma}{r}\right) u = -\kappa^2 u        \label{eq:A1}
\end{equation}
on $\RR^+$ such that $u(r)$ is locally square integrable
with measure $dr$ as $r\rightarrow \infty$, i.e., $u(r)\propto {\cal
W}_{|\gamma|/2\kappa; \ell+\frac{1}{2}}(2\kappa r)$. Clearly,
$h^\prime_\ell$ has eigenvalue $-\kappa^2$ if and only if
$u_\kappa(a) = 0$. Standard arguments give
\begin{equation}
u_{\kappa_2}(a)u^\prime_{\kappa_1}(a)-
u_{\kappa_1}(a)u^\prime_{\kappa_2}(a) = (\kappa_2^2-\kappa_1^2)
\int_a^\infty u_{\kappa_1}(r)u_{\kappa_2}(a) dr
\end{equation}
(the integral converges due to the boundary conditions on
$u_\kappa(r)$). From this it follows that firstly, $u_\kappa(a)$ has
isolated zeros as a function of $\kappa$, and secondly
\begin{equation}
\frac{d}{d (\kappa^2)} \left(
\left.\frac{u^\prime_\kappa(r)}{u_\kappa(r)}\right|_{r=a}\right)
= -\frac{\int_a^\infty u_\kappa(r)^2 dr}{u_\kappa(a)^2}.
\end{equation}
Hence $u^\prime_\kappa/u_\kappa |_{r=a}$ is increasing as a function
of $\kappa^{-1}$ between isolated poles, which correspond to
eigenvalues of $h^\prime_\ell$. Now write $v_\kappa$ for the solution
to~(\ref{eq:A1}) with regular boundary conditions at $r=0$. Clearly,
$\bar{h}_\ell$ has eigenvalues if and only if
$v^\prime_\kappa/v_\kappa |_{r=a}=u^\prime_\kappa/u_\kappa
|_{r=a}$. Similar arguments to those above show that
$v^\prime_\kappa/v_\kappa |_{r=a}$ is decreasing but everywhere
positive as a function of $\kappa^{-1}$. We conclude that between
every two eigenvalues of $\bar{h}_\ell$, there is exactly one pole
of $u^\prime_\kappa/u_\kappa |_{r=a}$, and hence exactly one
eigenvalue of $h^\prime_\ell$. Thus in particular, there are
infinitely many negative eigenvalues.

Next, we know from Lemma~\ref{Le:ineq} and Proposition~\ref{Pr:le}
that
$\mu_n(\bar{h}_\ell)=\nu_n(\bar{h}_\ell)\le\nu_n(h^\prime_\ell)$,
where $\nu_n(h^\prime_\ell)$ is determined by the min-max principle
(see Section 6). By the min-max principle, $\nu_n(h^\prime_\ell)$ is
either the $n$'th eigenvalue below the essential spectrum, or the
infimum of the essential spectrum. Hence $\inf\sigma_{\rm
ess}(h^\prime_\ell)\ge 0$. However, if $\nu_n(h^\prime_\ell)
=\inf\sigma_{\rm ess}(h^\prime_\ell)$ the min-max principle states
that there are at most $n-1$ eigenvalues below $\inf\sigma_{\rm
ess}(h^\prime_\ell)$. However, we have seen above that there are
infinitely many negative eigenvalues, so $\nu_n(h^\prime_\ell)$ is
the $n$'th eigenvalue $\mu_n(\bar{h}_\ell)$. Putting this with our
earlier observation shows that $\mu_n(h^\prime_\ell)\in
[\mu_n(\bar{h}_\ell),\mu_{n+1}(\bar{h}_\ell))$. That $\sigma_{\rm
ess}(h^\prime_\ell)=[0,\infty)$ may be seen by an explicit
consideration of the generalised eigenfunctions.

To prove the second statement, we define $r_L(\kappa)$ to be the
position of the first zero (away from $r=0$) of the generalised
eigenfunction of $h^L$ at energy $-\kappa^2$. By arguments similar to
the St\"{u}rm oscillation lemma one can see that $r_L(\kappa)$ is
increasing continuously in $\kappa$. Moreover, $r_0(0)=a_1$, so
$r_0(\kappa)>a$ for all $\kappa$. Inverting, we define $L(\kappa)$ to
be such that $r_{L(\kappa)}(\kappa)=a$. Then $L(\kappa)$ is
decreasing in $\kappa$ where it is defined, and, by the fitting
formula~(\ref{eq:fitc}), $L(0)=L_{\rm max}=\phi(a)/\theta(a)$, which
is strictly positive for $a<a_1$. Moreover, because $r_o(\kappa)>a$
for all $\kappa$, $L(\kappa)$ cannot vanish and so remains in
$(0,L_{\rm max}]$. It is clear that $h^\prime_0$ has an eigenvalue at
$-\kappa^2$ if and only if $h^{L(\kappa)}_0$ does, and also that
$\mu_n(h^\prime_0)=\mu_{n+1}(h^{L(\kappa)})$ because
$\mu_{n+1}(h^{L(\kappa)}_0)$ is the eigenvalue of $h^{L(\kappa)}$
between $\mu_n(\bar{h}_0)$ and $\mu_{n+1}(\bar{h}_0)$. But we also
know that $\mu_m(h^{L(\kappa)}_0)\le\mu_m(h^{L_{\rm max}}_0)$, so the
second statement is proved.

To prove the third statement, note that it is enough to determine
those values of $\kappa$ for which $u_\kappa(a)$ above vanishes.
Now $u_\kappa(a)$ may be written
\begin{equation}
u_\kappa(a) = e^{-\kappa a}(2\kappa a)^{\ell+1}
\Psi(\ell+1+\gamma/2\kappa, 2\ell+2;2\kappa a)
\end{equation}
where $\Psi(a,c;z)$ is the irregular confluent hypergeometric
function \cite{Grad}, so it is enough to find the zeros of
$\Psi(\ell+1+\gamma/2\kappa,
2\ell+2;2\kappa a)$. We look for solutions in the neighbourhood of
the Bohr levels $-\gamma/2\kappa = n+\ell$ for $n=1,2,\ldots$.
Writing $-\gamma/2\kappa = n + \ell + \xi$ and expanding in powers
of $\xi$, we find
\begin{eqnarray}
\Psi(\ell+1+\gamma/2\kappa, 2\ell+2;2\kappa a)  &=&
\Psi(1-n,2\ell+2;2(n+\ell)^{-1}a/a_0)  \nonumber \\
& & + K(\ell,n,a/a_0) \xi +O(\xi^2)
\end{eqnarray}
where
\begin{eqnarray}
K(\ell,n,\delta) & = & - \Psi^\prime(1-n,2\ell+2,
2\delta(n+\ell)^{-1}) \nonumber \\
& &
+2\delta\frac{(1-n)}{(n+\ell)^2}
\Psi(2-n,2\ell+3;2\delta(n+\ell)^{-1})
\end{eqnarray}
and $\Psi^\prime(a,c;z)\equiv (\partial/\partial
a)\Psi(a,c;z)$. Working to lowest order in $a/a_0$, we find
\begin{eqnarray}
\Psi(\ell+1+\gamma/2\kappa, 2\ell+2;2\kappa a) & \approx &
(-1)^{n-1}\frac{(n+2\ell)!}{(2\ell+1)!} \nonumber \\
 & & -\xi (-1)^{n-1}(n-1)!(2\ell)!
\left(\frac{2}{n+\ell}\frac{a}{a_0}\right)^{-(2\ell+1)}
\end{eqnarray}
and so we find the approximate solutions given in our third statement.

\newpage
{\large\bf Figure Captions:} \newline
\vspace{1.5in}

Figure 2.1: The function $G(z)$.
\end{document}